\documentclass[12pt]{paper}
\usepackage{lipsum}
\usepackage[T1]{fontenc}
\usepackage{textcomp, amssymb}
\usepackage{xcolor}
\usepackage{graphicx}
\usepackage{tabularx}
\usepackage{enumitem}
\usepackage{color,soul}
\usepackage{wrapfig}
\usepackage{comment}
\usepackage{setspace}
\usepackage{float}
\usepackage{environ}
\usepackage{lmodern}

\begin{document}
	
\noindent Title: In-situ X-ray analysis of misfit strain and curvature of bent polytypic GaAs-In$_x$Ga$_{1-x}$As core-shell nanowires
	
\noindent Mahmoud Al-Humaidi$^{1,2}$, Ludwig Feigl$^{2}$, Julian Jakob$^{2,3}$, Philipp Schroth$^{2,3}$, Ali AlHassan$^{2}$, Arman Davtyan$^{1}$, Jesús Herranz$^{4}$, Tasser Anjum$^{1}$, Dmitri Novikov$^{5}$, Lutz Geelhaar$^{4}$, Tilo Baumbach$^{2,3}$, Ullrich Pietsch$^{1}$	\\

\noindent $^1$ Solid State Physics, University of Siegen, Walter Flex Straße 3, 57068 Siegen.
\noindent $^2$ Institute for Photon Science and Synchrotron Radiation, Karlsruhe Institute of Technology, Hermann-von-Helmholtz-Platz 1, D-76344 Eggenstein-Leopoldshafen.
\noindent $^3$ Laboratory for Applications of Synchrotron Radiation, Karlsruhe Institute of Technology, Kaiserstraße 12, D-76131 Karlsruhe.
\noindent $^4$ Paul-Drude-Institut für Festkörperelektronik, Leibniz Institut im Forschungsverbund Berlin e.V., Hausvogteiplatz 5–7, 10117 Berlin.
\noindent $^5$ Deutsches Elektronen-Synchrotron, PETRA III, D-22607 Hamburg.

\noindent $^*$Email:mahmoud.humaidi@kit.edu

\begin{abstract}
Misfit strain in core-shell nanowires can be elastically released by nanowire bending in case of asymmetric shell growth around the nanowire core. In this work, we investigate the bending of GaAs nanowires during the asymmetric overgrowth by an In$_x$Ga$_{1-x}$As shell caused by avoiding substrate rotation. We observe that the nanowire bending direction depends on the nature of the substrate's oxide layer, demonstrated by Si substrates covered by native and thermal oxide layers. Further, we follow the bending evolution by time-resolved \textit{in-situ} X-ray diffraction measurements during the deposition of the asymmetric shell. The XRD measurements give insight into the temporal development of the strain as well as the bending evolution in the core-shell nanowire.
\end{abstract}

%
\vspace{2pc}

	\vspace{10pt}

\newpage
\section{Introduction} 
During the past decades, semiconductor nanowires have been studied intensively, because of their wide range of potential applications in modern and future electronics and optoelectronic devices  \cite{Berg2016,Gao2014,Tomioka2012,Bhuyan2018,Tatebayashi2015}. In particular, due to their high surface-to-volume ratio, nanowires can sustain strain more efficiently compared to their planar counterparts. 
A conventional way to induce strain in nanowires is the combination of materials with different lattice parameters in the form of axial or radial heterostructures \cite{Gronqvist2009,Johansson2011,Costas2020}. This strain modifies the optical properties of the respective device. For instance, growing an In$_{0.50}$Ga$_{0.50}$As shell around a thin lattice-mismatched GaAs nanowire core (with diameter < 40 nm) causes a 40\% reduction of the GaAs band-gap \cite{Balaghi2019}. For the case of uniform shell distribution, the induced strain as well as the decrease in the band-gap are laterally uniform, and the nanowire remains straight. The realization of a spatially varying strain field can be induced by bending the nanowire which opens up new possibilities for strain and band-gap engineering \cite{Lewis2018,Fu2014,Dietrich211}.
Spontaneous nanowire bending can be achieved by an asymmetric shell growth around a lattice mismatched nanowire core  \cite{Greenberg2019,Gagliano2018,Lewis2018}. The induced inhomogeneous strain distribution may cause the nanowire to bend fully backwards to contact either neighboring nanowires or the substrate serving as a new way to form electrical and optical interconnects \cite{Lewis2018}. Using molecular-beam epitaxy (MBE), the bending of the core-shell nanowires reported in \cite{Lewis2018} was achieved using a complicated growth scenario with substrate rotation and sequencing of group-III fluxes, to deposit the shell material on the same pre-define nanowire side resulting in nanowires bending to a defined direction. In other case, random nanowires bending along a non-defined direction was observed despite the fact that the template nanowires were rotating during lattice-mismatched shell growth \cite{Gagliano2018}. \\
In this research, we study the bending evolution of GaAs nanowires by growing an asymmetric In$_x$Ga$_{1-x}$As shell with $x= 15\%$ and $x=30\%$ nominal indium content using vapor-solid (VS) growth mode without performing substrate rotation. The growth was done by using a portable MBE (pMBE) chamber equipped with X-ray transparent beryllium windows. In order to trace the variation of the axial lattice parameter, we performed \textit{in-situ} XRD measurements on nanowires during shell deposition. These measurements give an access to the evolution of the strain and nanowire bending at the early stages which cannot be observed using the microscopic techniques, so far. \\
For this study we used samples of GaAs nanowire templates grown on Si(111) substrates covered by thin native oxide layer (< 2 nm) on one hand and thicker thermal oxide (10-20 nm), on the other hand. However, in contrast to the shell deposition method used by Lewis \textit{et al.} \cite{Lewis2018}, the entire process was carried out without sample rotation in order to maintain the consistent X-ray diffraction condition. We observed that nanowires grown on Si substrate with native oxide bend away from the group-V (As) flux whereas nanowires grown on Si with thermal oxide bend away from group-III (Ga) flux. This difference in the bending direction indicates a significant influence of the substrate oxide type, in addition to flux directions \cite{VanTreeck2019}, to the preferable nanowire side facets for VS shell deposition. By means of scanning electron microscopy (SEM) we show  that, despite of the different bending directions, the nanowire curvature is comparable for both oxide types using similar MBE growth conditions.
\noindent These findings are confirmed by time-resolved \textit{in-situ} XRD measurements carried out on nanowire ensembles during shell deposition. Furthermore, the nanowire axial strain and curvature as functions of shell growth time show a nonlinear dependency at the early shell growth stages. \\
Finally, we extend our study by performing {\textit{in-situ}} XRD measurement on a single nanowire during the shell growth. The observed dependency of the single nanowire bending on shell growth time with the corresponding shell thickness is in agreement with the findings obtained from the nanowire ensembles.

\section{Experimental  details}
\subsection{Sample preparation}

In our study, we used n-doped Si(111) substrates covered by  naturally grown Silicon oxide (referred to as “native oxide” in the text) and thicker thermally-grown oxide layer (referred to as “thermal oxide” in the text) with patterns of drilled nano-holes. The pattern structure is designed in such a way that we can investigate an ensemble of nanowires simultaneously, on one hand, and an individual nanowire, one the other hand. The two types of substrates require different sample treatments. The native oxide substrates pass through regular solvent cleaning (two rounds of dipping into Acetone, Isopropanol and Ultra-pure water baths in ultrasonic cleaner respectively). The patterned substrates with thermal oxide require Hydrofluoric (HF) acid etching following the same regular solvent cleaning used for native oxide substrates. The etching was done using 0.5\% HF for 1 minute to reduce the thickness of the naturally grown oxide on the substrate surface within the nano-holes. After oxide etching, the thermal oxide substrates were dipped into boiled ultra-pure water for 10 minutes for surface smoothing \cite{Kupers2017}. Afterwards, the substrates were degassed under the ultra-high vacuum condition at 300$^{\circ}$C for 30 min to get rid of solvent residuals before loading them into the MBE growth chamber.
We investigated five samples in this study as listed in table \ref{tab1} where the mean nanowire length ($l_{NW}$), shell thickness approximated from the calibration of the shell material growth in 2D planar system and the In concentration are listed there as well (the samples are numbered according to the performed study).

\noindent The growth of all samples was performed in a special portable molecular beam epitaxy (pMBE) chamber \cite{Slobodskyy2012}, which is designed for \textit{in-situ} X-ray characterization experiments at synchrotron radiation facilities. The pMBE chamber is equipped with solid source effusion cells of Ga and In, and valved cracker cell supplying As$_{4}$. All cells are inclined to the substrate normal by $\Phi=28^{\circ}$. The azimuthal angles are 60$^{\circ}$ between the Ga and the In cells and 120$^{o}$ between the Ga and the As$_{4}$ cells (see figure \ref{fig:1_2}(a)). 

\begin{table}
	\centering
		\begin{center}
	\caption{\label{tab1} List of the investigated samples and the main properties required for this research.}
	\footnotesize
		\begin{tabular}{lccccc}

		Sample      &     Oxide type & Experimental  & Nanowire            & Shell                & In content    	\\
	                &     & technique     &   length(nm) ±100  & growth time (minute)   &   (\%)	        \\
\hline

		Sample 1    &	Native oxide  &	SEM           &  1800	            &30	                   & 30             \\
		Sample 2    &	Thermal oxide &	SEM	          &  1800	            &30	                   & 30             \\
		Sample 3    &	Native oxide  &	XRD+SEM	      &  1200           	&20                    & 15             \\
		Sample 4    &	Thermal oxide &	XRD+SEM	      &  1200	            &20                    & 15             \\
		Sample 5    &	Thermal oxide & XRD+SEM	      &  1200	            &11                    & 15             \\

\end{tabular}
		\end{center}
\end{table}
\normalsize

\noindent The Si substrates of samples 2, 4 and 5 were covered by a 15-20 nm thick thermal oxide layer patterned with nano-hole arrays defined by electron-beam lithography. These nano-holes act as nucleation sites for Ga catalyst droplets for the epitaxial nanowire growth. The pattern consists of several equidistant $100\times100 \: \mathrm{\mu m^2}$ large arrays of nano-holes arranged in a hexagonal grid. The separation between neighboring holes (pitch $p$) differs for each array, ranging from $p=0.1 \: \mathrm{\mu m}$ to $p=10 \: \mathrm{\mu m}$. Apart from the ensemble arrays, single nano-holes with $10$ $\mathrm{\mu m}$ separation are drilled along a $1\: \mathrm{mm}$ straight line bordered by two etched markers to facilitate the illumination of single nanowire (see figure \ref{fig:S1_2}(i) in the supplementary materials). More details about substrate preparation can be found in \cite{Kupers2017}.
The sample layout enables the navigation on the substrate during the XRD measurements facilitating access to nanowire arrays (samples 2 and 4) and single nanowire (sample 5). The recipe to find a single nanowire using this substrate layout can be found in \cite{AlHassan2018}. In this study we investigate the nanowires in arrays with pitch size $p=1 \: \mathrm{\mu m}$ in order to avoid the flux shadowing effect by the neighboring nanowires in case of lower pitch size \cite{Schroth2018}. \\
Material fluxes were calibrated via the respective layer-by-layer growth on GaAs (001) substrate monitored by reflection high-energy electron diffraction (RHEED). The readout of the substrate temperature was calibrated to the temperature of Ga-oxide removal on epi-ready GaAs(001). Within the holes Ga catalyst droplets were formed at T$_{sub}$ {$\approx$} 630$^{\circ}$C by depositing Ga with an equivalent GaAs thickness of 50 monolayer (ML). Epitaxial VLS growth of the GaAs nanowire cores was initiated by supplying As and Ga simultaneously at T$_{sub}$ {$\approx$} 630$^{\circ}$C and using V/III ratio of 3 and a growth rate of 0.12 ML/s. 
The growth of the GaAs nanowires proceeded for 45 minutes for sample 2 and 30 minutes for samples 4 and 5 resulting in different mean nanowire lengths ($l_{NW}$)(see table \ref{tab1}).\\ 
For samples 1 and 3 which were grown on native oxide substrates, prior to nanowire growth procedure used for the other samples, Ga droplets were formed at T$_{sub}${$\approx$}600$^{\circ}$C followed by annealing step for 15 minutes at T$_{sub}${$\approx$} 800$^{\circ}$C to evaporate the Ga droplets completely from the substrate surface. Randomly distributed nano-holes (i.e. nucleation sites) on the native oxide result from this procedure \cite{Tauchnitz2017}. The growth time for the GaAs nanowires was 45 minutes for sample 1 and 30 minutes for sample 3 (see table \ref{tab1}). \\
To initiate the VS shell growth for all samples, first the Ga droplets at the nanowire apex were consumed to prevent any further axial VLS-growth. In the case of self-catalyzed GaAs nanowires, this is done by terminating the Ga supply while continuously supplying As at growth temperature. Afterwards, the substrate temperature and the material fluxes were adjusted to the desired parameters. The reduced substrate temperature T$_{sub}$ (400$^{\circ}$C (for samples 1-4)–  500$^{\circ}$C (for sample 5)) and  the higher V/III ratio (V/III{$\approx$} 4(for sample 5) – 6 (for samples 1-4)) reduce the diffusion length of the III-material favoring VS growth on the nanowire sidewalls.\\
The shell growth rate was chosen to be 0.04 ML/s which is equivalent to 0.8 nm/min for samples 1-4 and 0.25 ML/s (0.5 nm/min) for sample 5. The shell growth time was 30 minutes for samples 1 and 2, 20 minutes for sample 3, 25 minutes for sample 4 and 11 minutes for sample 5. The resulting shell thicknesses are listed in table \ref{tab1} complemented by SEM inspection, where the shell thickness was evaluated in addition to the calibration method by comparing the final nanowire diameter with the GaAs nanowire diameter of respective reference samples.

\subsection{X-ray diffraction measurements}
XRD serves as an ideal technique to investigate nanowire bending due to its high sensitivity to small changes in the crystal structure and orientation whereas these changes are indicated by the changes of the diffraction signal profile. Therefore, \textit{in-situ} XRD measurements were performed on GaAs nanowires during In$_{0.15}$Ga$_{0.85}$As shell growth. The experiments were carried out at the German Electron Synchrotron (DESY-Hamburg). 
The {\textit{in-situ}} XRD measurements on sample 2 and sample 5 were performed at the Resonant Scattering and Diffraction beamline P09 \cite{Strempfer2013a} whereas nanowire arrays of sample 4 were measured at the \textit{In-situ} and Nano-X-ray diffraction beamline P23 \cite{DESY}. Both beamlines are equipped with a heavy load goniometer which can withstand the weight of the pMBE. The beam was focused to a spot size of few microns (1.5 $\mathrm{\mu m}$  horizontal  $\times$ 5 $\mathrm{\mu m}$ vertical) by means of compound refractive lenses which is essential to illuminate only the desired regions on the substrate. During all {\textit{in-situ}} XRD experiments, 2D pixel detectors were used for recording the diffracted signals.
We measured reciprocal space maps (RSMs) in the vicinity of the GaAs(111) Bragg reflection in order to identify the bending direction in addition to measuring the nanowire bending angle (thus nanowire curvature) and the axial strain induced by the lattice-mismatched shell. Choosing a beam energy of 15 keV, the incident and scattering angles of the X-ray beam for the GaAs(111) Bragg reflection are 7.27$^{\circ}$ and 14.54$^{\circ}$, respectively. These values are smaller than the maximum opening angle defined by the size of pMBE beryllium windows. \\
Rocking the GaAs(111) lattice planes around the Bragg angle of the respective reflection, the 3D intensity distribution of the Bragg reflection was recorded as a function of the incident angle and the horizontal and vertical angular ranges covered by the 2D detector. The angular coordinates are then translated to reciprocal space vectors using the equations listed in \cite{U.PietschT.Baumbach2004}. These vectors are $Q_x$, $Q_y$ and $Q_z$ and represent the 3D RSMs. Here $Q_z$ is set parallel to the GaAs[111] nanowire growth axis and is sensitive to the variation in the axial lattice spacing ($d_{111}$), i.e the axial strain $\epsilon_{||}$. The orthogonal components $Q_x$ and $Q_y$, set parallel to the plane of the Si(111) substrate ($Q_x$ is parallel to [0.11] and $Q_y$ is parallel to [211]), are sensitive to changes in the crystal orientation (i.e. tilting and bending of the nanowire).\\
For samples 2 and 4, the shell growth and the XRD measurements were done in multiple cycles. At the end of each shell growth run, a RSM of the full 3D GaAs(111) Bragg reflection was recorded to track the evolution of the diffraction peaks caused by nanowire bending. 
For the single nanowire measurement (sample 5), we considered only few selected 2D cuts through the 3D RSM to improve the time resolution of monitoring the XRD signal during nanowire bending. \\
The samples measured by XRD were intended to show a smaller bending compared to the samples inspected only by SEM because the diffraction signal could be followed only up to the maximum diffraction angle defined by the size of the pMBE beryllium windows. Therefore, the XRD data refer rather to the early stages of shell growth and nanowire bending.\\

\section{Results and discussion}
\subsection{Bending direction}

\begin{figure}[!ht]
 \centering
	\includegraphics[width=\textwidth]{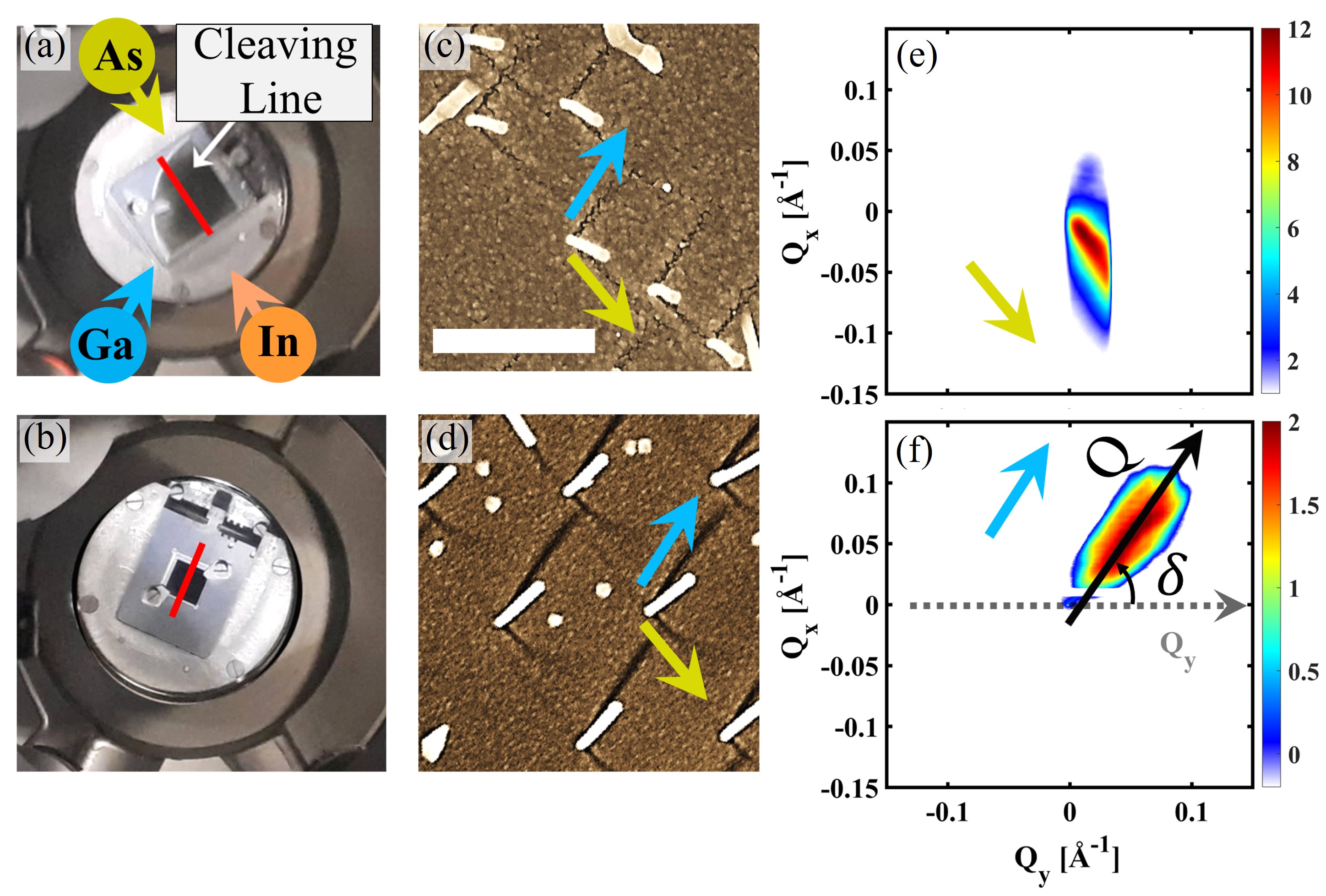}
	
	\caption{(a) and (b) Pictures of the geometrical arrangement of the growth material cells of the pMBE and the orientation of the quarter-inch substrate with native oxide layer (sample 1) and $1\times 1$ cm patterned substrate with thermal oxide layer (sample 2), respectively. (c) and (d) Top-view SEM images of the bent nanowires of sample 1 and sample 2, respectively (The scale bar corresponds to $1\mu$m). (e) and (f) Projection of GaAs(111) Bragg reflection on the ($Q_{x}Q_{y}$) plane of RSM of the bent nanowires grown on substrate with native oxide layer (Sample-3) and substrate with thermal oxide layer (sample 4) recorded after shell growth, respectively.}
	\label{fig:1_2}
\end{figure}

Since the substrate within the MBE during the growth is stationary (i.e. the alignment of the nanowire templates with respect to the MBE cells is fixed), the In$_{x}$Ga$_{1-x}$As shell materials are deposited mainly on the nanowire side walls that face the material fluxes. The lattice parameters of the shell material is larger than that ones of the core, therefore, the induced mismatch axial strain (which is parallel to the nanowire axis $\epsilon_{||}$) causes the nanowire to bend away from the side at which the shell is deposited. The orientation of the material sources with respect to the Si substrate are shown in figure \ref{fig:1_2}(a). For sample 1 (native oxide) we aligned the substrate in a way that a defined substrate edge is parallel to the direction of As flux (figure \ref{fig:1_2}(a)). This edge is used as a reference for the later cleaving of the substrate (the red line in figure \ref{fig:1_2}(a) indicate the cleaving edge). On the other hand, the thermal oxide substrate (sample 2) is aligned such that the cleaving edge directs to the Ga flux shown in figure \ref{fig:1_2}(b). This alignment enables a directional reference for the SEM inspection and side-view SEM evaluation of nanowire curvature. In top-view SEM, the bent nanowires appear as elongated objects and thus allow the extraction the bending direction, thanks to our proper sample alignment in the chamber. As shown in figures \ref{fig:1_2}(c) and \ref{fig:1_2}(d) the bending direction differ for both samples.\\
In case of sample 1, the nanowires bend away from the As flux (yellow arrow in figure \ref{fig:1_2}(c)), whereas in case of  sample 2, the nanowires bend away from the Ga flux (blue arrow in \ref{fig:1_2}.d). This is an indication, that the position of the In$_x$Ga$_{1-x}$As shell in case of the thermal oxide (sample 2) is given by the position of the Ga-source, whereas it is given by the position of the As-source in case of the native oxide (sample 1). \\
During the XRD measurements, several hundreds of nanowires were illuminated at the same time giving their mean properties and mean geometrical alignment. Using the same experimental conditions, the relative orientation of the nanowires geometry with respect to the reciprocal space geometry is consistent for all samples. Figures \ref{fig:1_2}(e) and \ref{fig:1_2}(f) show $Q_{x}Q_{y}$ projection of the 3D RSMs of the GaAs(111) reflection for samples 3 and 4. The elongation of the Bragg peak clearly evidences that the bending directions are homogeneous within one substrate, but different for the two oxide types. \\

\begin{figure}[!ht]
	\centering
	\includegraphics[width=\textwidth]{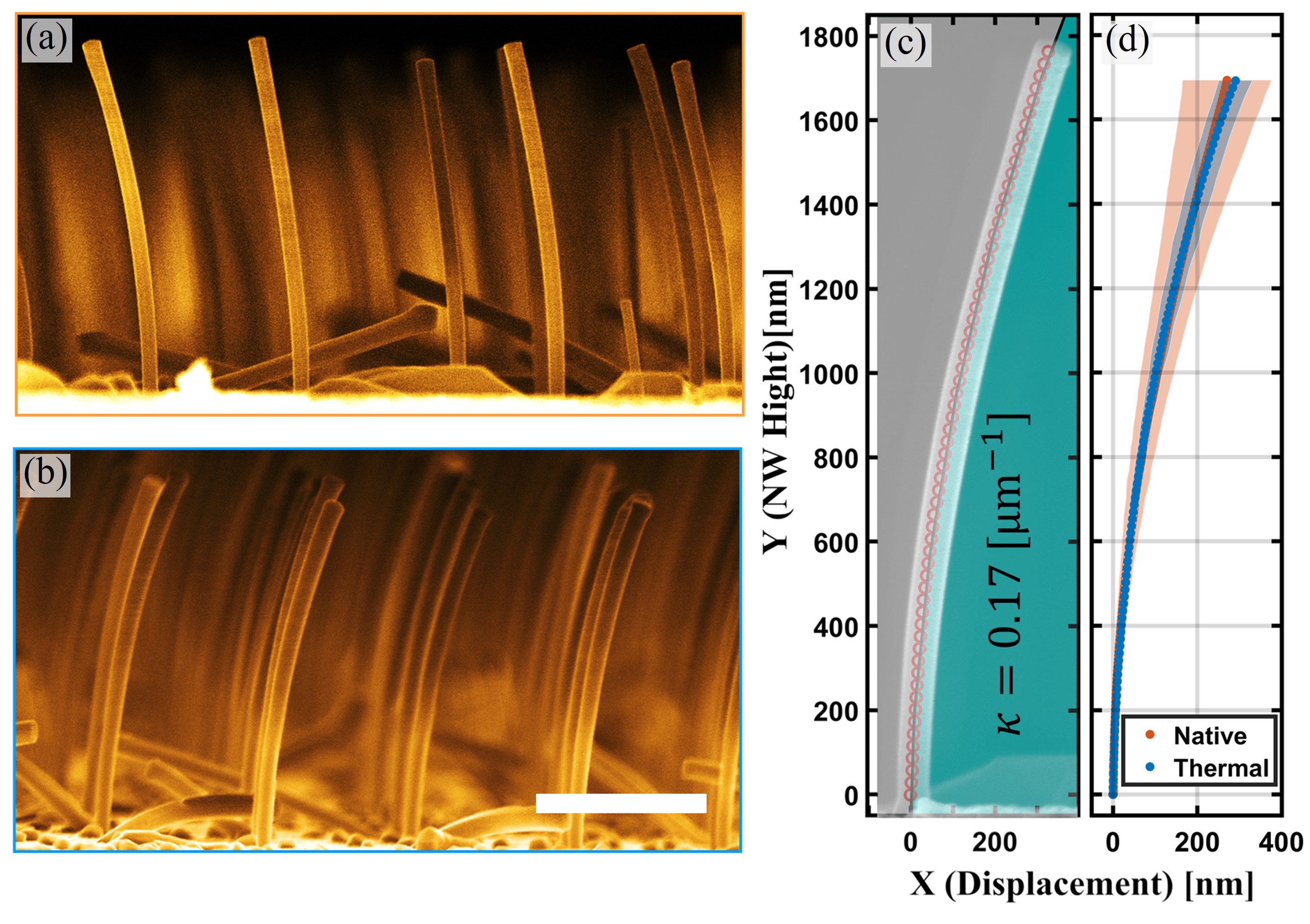}
	\caption{(a) and (b) side-view SEM images of bent nanowires grown on Si substrate with native oxide (sample 1) and thermal oxide (sample 2), respectively (The scale bar corresponds to 1$\mu$m). (c) Side-view SEM image of selected bent nanowire with extracted curvature profile and a circle fitting function for exemplary illustration of the SEM evaluation method of nanowires curvature. (d) Comparison of the curvature profile of the nanowires of the mentioned samples, where the shaded areas indicate the curvature variation and the lines are their mean values.}
	\label{fig:2_2}
\end{figure}

\noindent The careful alignment of the samples during growth further made it possible to cut the samples in such a way that side-view SEM investigations could be performed perpendicular to the bending directions. Figures \ref{fig:2_2}(a) and \ref{fig:2_2}(b) show side-view SEM images of bent core-shell nanowires grown on native oxide (sample 1) and thermal oxide (sample 2), respectively. The side-view images reveal that the curvature of the nanowires in sample 2 is more uniform than that of nanowires in sample 1. For quantitative analysis of the curvature, 47 nanowires from sample 1 and 40 nanowires from sample 2 were analyzed. Several data points were selected along the nanowire axis to obtain the bending profile. An exemplary visualization of this approach is demonstrated in figure \ref{fig:2_2}(c). The curvature $\kappa$ of this nanowire is about 0.17 $\mathrm{\mu m^{-1}}$ which was determined by fitting its profile using a circle function.\\ 
In figure \ref{fig:2_2}(d) the mean nanowire bending profiles of both samples (i.e. samples 1 and 2) are plotted. Both samples have the same mean curvature as shown in red and blue lines, respectively.
However, the local variation of the nanowire curvature is larger in case of native oxide substrates as shown by red and blue shades in figure \ref{fig:2_2}(d). This finding can be explained by a larger variation of the GaAs nanowire diameters when grown on native oxide substrates. The overall evaluation proves the homogeneous curvature along the nanowire full length, which implies axial-homogeneity of shell deposition along the nanowire growth axis.

\subsection{Bending Evolution during shell growth}

\begin{figure}[!ht]
	\centering
	\includegraphics[width=\textwidth]{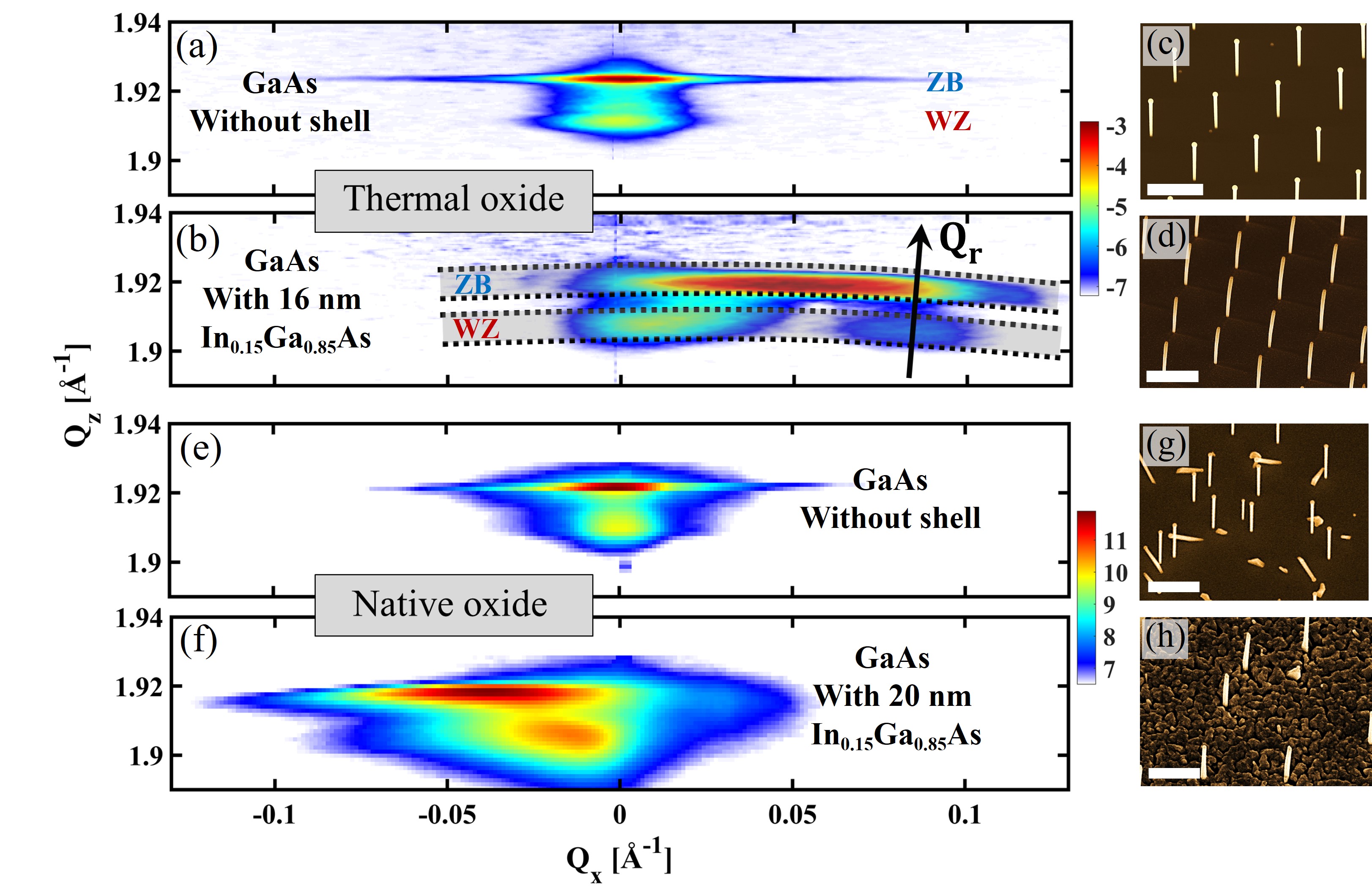}
	\caption{(a) and (b) GaAs(111) Bragg's reflection from nanowire ensembles on thermal oxide substrate (sample 4) projected on ($Q_xQ_z$) plane of RSMs recorded before and after the deposition of about 16 nm thick In$_{0.15}$Ga$_{0.85}$As shell, respectively. (c) SEM images with 30$^{o}$ tilt-view of GaAs nanowire of a reference sample grown under same growth condition. (d) SEM images with 30$^{o}$ tilt-view of sample 4. (e) and (f) GaAs(111) Bragg's reflection from nanowire ensembles on native oxide substrate (sample 3) projected on ($Q_xQ_z$) plane. (g) and (h) are the respective 30$^{o}$ tilt-view SEM images, where (g) is from a reference sample and (h) from sample 3. (All scale bars correspond to $1\mu m$).}
	\label{fig:3_2}
\end{figure}

The analysis of the temporal evolution of the bending process and the axial strain was performed by {\textit{in-situ}} XRD experiments. The axial strain and bending can be distinguished by the changes of Bragg peak profile in the RSMs (see figure \ref{fig:3_2}) as follows:
\begin{itemize}
\item Nanowire bending results in continuous change in the orientation of the GaAs(111) lattice planes resembled by broadening of the Bragg peak along a sphere in the RSM (see figures \ref{fig:1_2}(e), \ref{fig:1_2}(f), \ref{fig:3_2}(b), \ref{fig:3_2}(f)). 

\item In case of no axial strain, the RSM sphere has a radius equals to the amplitude of $Q_{z}$ which gives the inter-planar spacing ($d_{111}$) of the unstained nanowire crystal. 

\item When the axial strain is induced, the radius of the RSM sphere changes accordingly. In case of the tensile strain in the nanowire core (i.e. increment of $d_{111}$ of GaAs crystal) the XRD signal shifts toward lower $Q_{z}$ value (thus smaller radius of the RSM sphere). The broadening of the peak along $Q_r$ as highlighted with gray arcs in figure \ref{fig:3_2}(b) results from continues changes of the lattice parameter across the nanowire cross section (i.e. gradient strain).  

\end{itemize}

To evaluate the spatial changes of the signal on the RSM sphere, we introduce two new reciprocal space vectors $Q$ and $Q_r$ which share the same origin as $Q_x$, $Q_y$ and $Q_z$. The first vector $Q$ is defined along the elongation of Bragg reflection as indicated by black arrow in figure \ref{fig:1_2}(f). The amplitude of $Q$ follows as $Q=\frac{|Q_y|}{cos(\delta)}$ where $\delta=56^{\circ}$ is the angle between the Bragg peak elongation direction (i.e. bending direction) and $Q_y$. The other vector $Q_r=Q_z\:sin(\beta)$ is tilted from $Q_z$ by the bending angle $\beta$ of the nanowire as indicated by black arrows in figure \ref{fig:3_2}(b) (the illustration of the 3D arrangement of all RSM vectors can be seen in figure \ref{fig:S3_2}). Therefore, to measure the axial strain induced by the shell in the nanowire, the XRD signal must be evaluated along $Q_r$ in the RSMs. \\
\noindent In order to record the whole nanowire bending, one needs to scan a wider range in reciprocal space compared to straight nanowires in order to cover the full elongation of the Bragg peak. The polytypism of GaAs nanowires can be utilized for a more accurate evaluation of the nanowire bending angle. In the first stages of Ga-assisted GaAs nanowire growth, the Ga droplet is unstable and small, resulting in the preferential growth of the WZ phase at the bottom of the nanowire (denoted by WZ$_{Bottom}$) and the inclusion of stacking faults \cite{Shtrikman2009}. During subsequent growth, the ZB phase and its rotational twin (TZB) are formed \cite{Rieger2013,Yamashita2010}. To initiate radial VS growth, the axial VLS growth needs to be terminated by consuming the Ga catalyst. The consumption of the droplet changes the growth conditions and faulted ZB segments and WZ segments are formed on the top of the nanowire (denoted by WZ$_{Top}$) \cite{Jacobsson2016,Schroth2018}. Notably, when measuring the GaAs(111) Bragg reflection, ZB and its twin overlap at the same position in reciprocal space, and therefore cannot be distinguished, whereas the WZ is separated from the ZB signal.\\
Figures \ref{fig:3_2}(a) and \ref{fig:3_2}(b) show $Q_xQ_z$ projections of the GaAs(111) Bragg reflection of nanowires grown on the thermal oxide substrate (sample 4) before and after shell growth, respectively, indicating nanowire bending. In both figures, the XRD peaks of ZB and WZ are clearly discernible. Comparing figures \ref{fig:3_2}(a) and \ref{fig:3_2}(b), the WZ and ZB peaks show significant changes, i.e.  elongation of the ZB peak, and elongation and splitting of the WZ peak. SEM images with 30$^{\circ}$ tilt-view of the straight and bent nanowires are shown in figures \ref{fig:3_2}(c) and \ref{fig:3_2}(d), where the straight GaAs nanowires are from a reference sample grown under identical growth conditions of the studied sample.

\noindent To measure the mean curvature of the illuminated nanowires, one has to evaluate the elongation of the Bragg peaks along $Q$. To do so, we integrate the ZB and WZ signals on the RSM spheres with radii equal to the amplitude of $Q_r$  for each polytype.
Similar {\textit{in-situ}} XRD measurement was performed on nanowires grown on native oxide substrate (sample 3). The XRD signals of these wires before and after shell growth are shown in figures \ref{fig:3_2}(e) and \ref{fig:3_2}(f). 
The different elongation directions of the (111) Bragg peak for nanowires grown on thermal oxide substrate of sample 4 (figure \ref{fig:3_2}(b)) and native oxide substrates of sample 3 (figure \ref{fig:3_2}(f)) confirm the dependency of the bending direction on the oxide type. In further XRD study we focus on the nanowires grown on thermal oxide substrates (i.e. samples 4 and 5).

\begin{figure}[!ht]
	\centering
	\includegraphics[width=\textwidth]{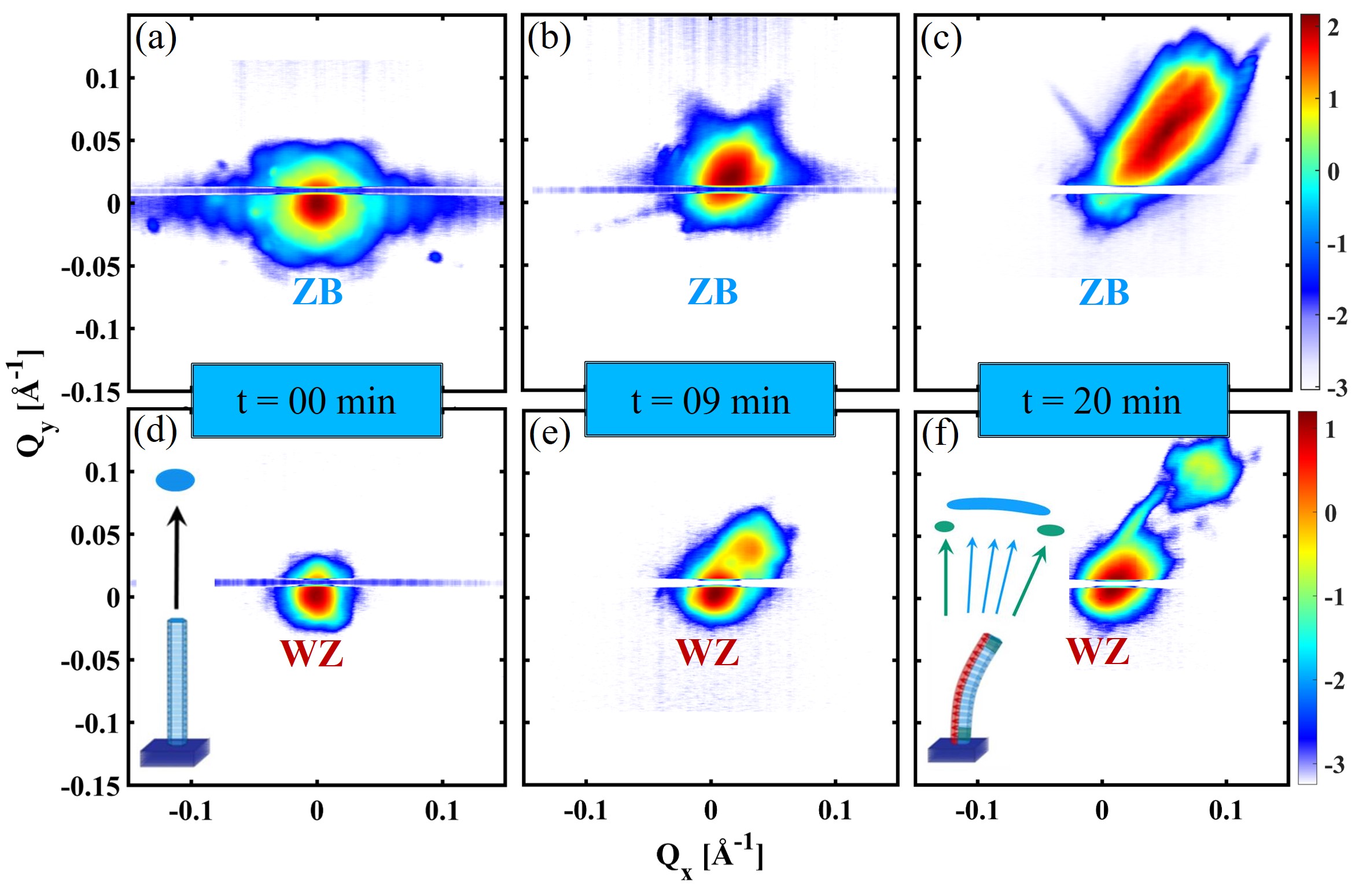}
	\caption{\textit{In-situ} XRD measurement of sample 4. (a), (b) and (c) GaAs(ZB) Bragg’s reflections before shell growth, after 9 minutes and 20 minutes of shell growth, respectively, (d), (e) and (f) are the  GaAs(WZ) Bragg’s reflections at the same time stages.}
	\label{fig:4_2}
\end{figure}

\noindent For nanowire that is straight and vertical to the substrate surface, all ZB lattice planes are flat and parallel in real space, which results in a sharp XRD peak at $(Q_x,Q_y)=0$ \AA$^{-1}$ (i.e. $Q=0 $ \AA$^{-1}$) in RSM (figure \ref{fig:4_2}(a)) with a width corresponds to the divergence of the probing X-ray beam. As soon as the nanowire starts to bend, the orientation of the lattice planes becomes a function of their position along the nanowire axis in real space causing broadening of the XRD peak in RSM. Figures \ref{fig:4_2}(b) and \ref{fig:4_2}(c) show the elongation of the XRD signal in $Q_xQ_y$ at different stages during shell growth time (before shell growth, after 9 minutes and after 20 minutes of shell growth). By estimating the integrated intensity ratio of the ZB peak to the total signal, it is apparent that the ZB phase represents about 80$\%$ of the crystal structure of the nanowire.\\

\noindent However, the XRD signals of the WZ phase behave differently from the signal of ZB due to their distribution along the nanowire where WZ phase locates mainly at the bottom and the upper part of the nanowire. In straight nanowires, the two WZ segments (i.e. WZ$_{Bottom}$ and WZ$_{Top}$) are located at the same position at $(Q_x,Q_y)=0$ \AA$^{-1}$ in reciprocal space (\ref{fig:4_2}.d). As soon as the nanowire starts to bend, the WZ$_{Top}$ peak moves accordingly, whereas the WZ$_{Bottom}$ peak changes by negligible amount. This leads to peak splitting in reciprocal space (figures \ref{fig:4_2}.e and \ref{fig:4_2}.f). \\
Therefore, by knowing the mean length of the nanowires, the complete circular bending of the nanowire can be extracted from the separation of the two WZ peaks where this methods provide high accuracy.

\begin{figure}[!ht]
	\centering
	\includegraphics[width=\textwidth]{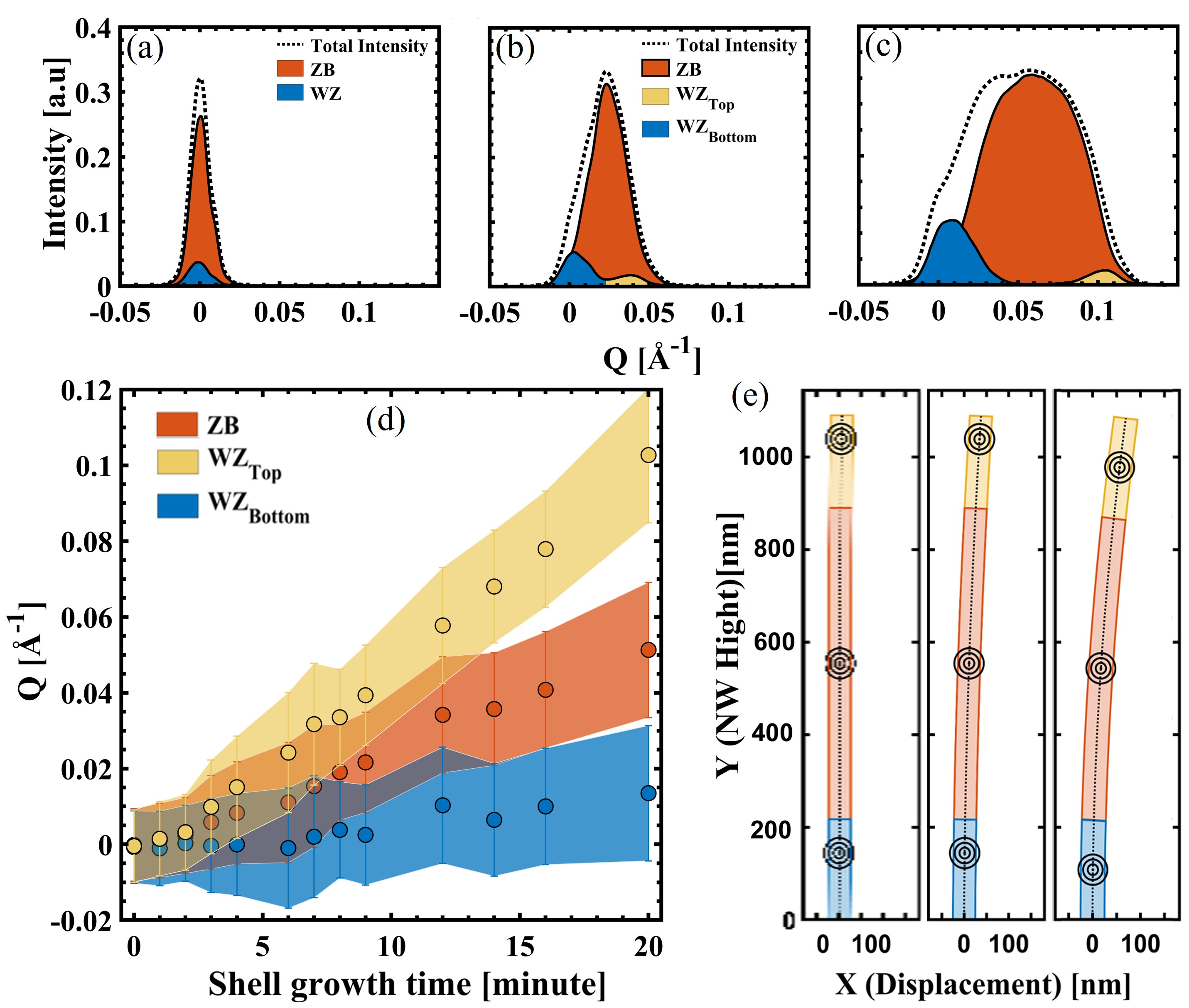}
	\caption{Profiling the XRD signals of sample 4. (a), (b) and (c) are intensity profiles of the XRD signals of the respected polytypes at the three different growth stages as mentioned in figure \ref{fig:4_2}(a-f) integrated along $Q$. (d) The peak positions of ZB, WZ$_{Top}$ and WZ$_{Bottom}$ along $Q$ plotted as function of shell growth time. (e) Modeled phase stacking based on circular bending.}
	\label{fig:5_2}
\end{figure}

\noindent The integrated intensities along $Q$ and the resulting line profiles are represented in figures \ref{fig:5_2}(a)-\ref{fig:5_2}(c) for the three different growth stages demonstrated in figure \ref{fig:4_2} (i.e. before shell growth, after 9 minutes and 20 minutes of shell growth). Before shell growth, the WZ (shaded in blue) and ZB (shaded in orange) peaks are both located at $Q = 0 $ \AA$^{-1}$. After 9 min of shell growth, the WZ peak splits into two separate ones. One peak remains close to $Q = 0 $ \AA$^{-1}$ with slight broadening towards higher $Q$ values (shaded in blue in figure \ref{fig:5_2}(b) and \ref{fig:5_2}(c)), which we attribute to the WZ$_{Bottom}$ segment. The second peak, centered at higher $Q$ values, (shaded with yellow in figure \ref{fig:5_2}(b) and \ref{fig:5_2}(c)), originates from the WZ$_{Top}$ segment. 
After 20 min of shell growth, the broadening of the ZB peak along $Q$ figure \ref{fig:5_2}(c) and the separation between the two WZ peaks figure \ref{fig:5_2}(f) increases further. The center of ZB peak during shell growth is located in between the two WZ peaks confirming the phase distribution along the nanowire. 
To track the evolution of the ZB and the two WZ peaks along $Q$ as a function of shell growth time, the peak centers of the respective polytypes were extracted by fitting their XRD line profiles using Gaussian function, resulting in the plots in figure \ref{fig:5_2}(d) (2D cuts of RSMs on $Q$ and $Q_z$ and the extracted line profiles along $Q$ and $Q_r$ are shown in figure \ref{fig:S2_2}). 
As the bending is found to be homogeneous the curvature $\kappa$ of the bent nanowires can be described by a specific bending radius ($r= \kappa^{-1}$) and the bending angle $\beta$ with respect to the normal of the substrate. The two parameters are related by $\beta= \kappa \: l_{NW}$ where $l_{NW}$ denotes the nanowire length measured by SEM. The bending angle $\beta$ is determined from the shifting of $WZ_{top}$ XRD signal along $Q$ using the basic relation $\tan(\beta)=\frac{Q(WZ_{Top})}{Q_{r})}$ (as demonstrated in figure \ref{fig:S3_2}).\\
\noindent The tilt angle of (WZ$_{Top}$) at the final stage (i.e. after 20 min of shell growth which corresponds to 16 nm shell thickness) is $\beta\approx3^{\circ}$ considering the measured average length of the nanowires of $l_{NW} = 1100$ nm. In this case the resulting final curvature is $\kappa\approx0.047 \: \mathrm{\mu m^{-1}} $. The distribution of both polytypes along the nanowire as well as the calculated nanowire bending at the three selected growth stages by our model of circular bending are visualized in figure \ref{fig:5_2}(e).

\begin{figure}[!ht]
	\includegraphics[width=\textwidth]{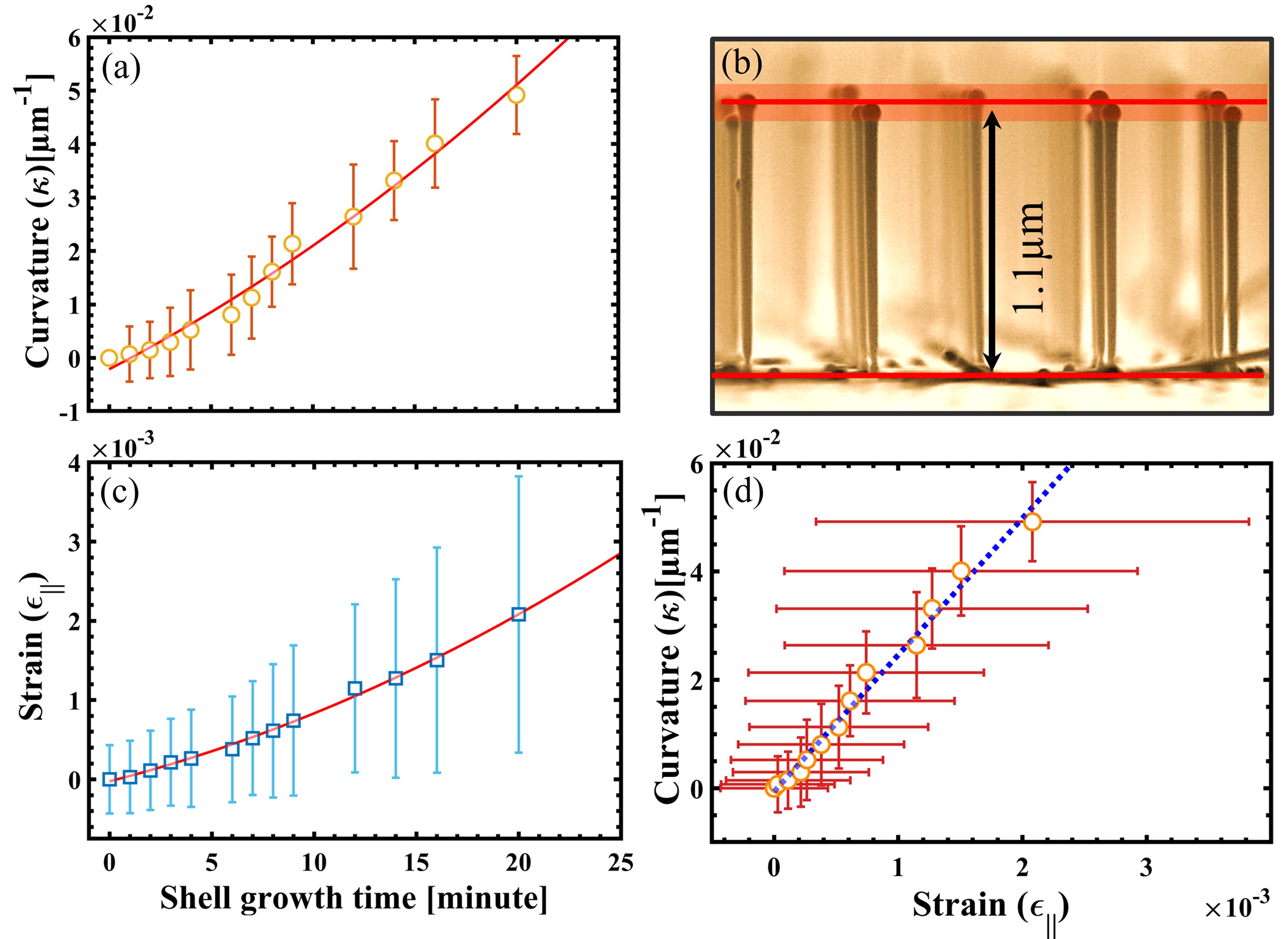}
	\caption{(a) Evolution of nanowire curvature as function of shell growth time. The curvature is measured by considering the nanowire length of a reference sample grown under identical growth conditions of the sample shown in (b). (c) Strain evolution in the ZB polytype as function of shell growth time. (d) Evolution of nanowire curvature as function of induced strain extracted from the \textit{in-situ} XRD measurement of sample 4. }
	\label{fig:6_2}
\end{figure}

\noindent By measuring the mean nanowire bending angle as function of shell growth time, the evolution of nanowire curvature during shell growth is plotted as function of shell growth time in figure \ref{fig:6_2}(a). The curvature was determined by considering the mean nanowire length from a reference sample of straight nanowires shown in figure \ref{fig:6_2}(b). For strain analysis, we consider only the strain induced in the GaAs(ZB) phase, since it is the prominent structure along the nanowire. The axial strain is extracted from the shift of the ZB peak along $Q_r$ and plotted as function of shell growth time where the broadening of the peaks was considered to estimate the strain variation as indicated by the bars as shown in figure \ref{fig:6_2}(c) (see the supplementary materials). \\
From figures \ref{fig:6_2}(a) and \ref{fig:6_2}(c), it is evident that the evolution of the nanowire curvature and the strain are not linear with the shell growth time. During the first shell growth run of up to around 5 minutes of shell growth, minor curvature and strain occur. Exceeding the growth time of 5 minutes, the changes in nanowire curvature increases and reaches a value of about $\kappa\approx0.05 \: \mathrm{\mu m^{-1}}$ after 20 minute of shell growth where the axial strain reaches a value of $\epsilon_{||}\approx0.003$.\\ 
Nevertheless, the relation between the curvature and strain is linear as shown in figure \ref{fig:6_2}(d) which implies that the linear elasticity of the studied core-shell system is still valid. The non-linearity of relation between the shell growth time and the evolution of the strain and the curvature may be attributed to the changes of the growth dynamics during shell growth.\\

The axial lattice spacing $d_{111}$ of the un-strained ZB can be measured from the diffraction peak on $Q_r$ by $d_{111}(GaAs(ZB))=\frac{2\pi}{Q_{r}(GaAs(ZB))}$.

\noindent The position of the Bragg peak on $Q_r$ of the un-strained ZB at growth temperature is $Q_{r}(GaAs(ZB))=1.9239$ \AA$^{-1}$ therefore $d_{111}(GaAs(ZB))=3.2659$ \AA and the corresponding lattice parameter $a_{(GaAs(ZB))}=5.6566$ \AA.

The measurement was performed at growth temperature and the thermal expansion of the GaAs crystal is considered. However, using Vegard’s law \cite{Denton1991}, the lattice parameter of the shell is $a_{(In_{0.15}Ga_{0.85}As)}=5.7174$ \AA  and which is the maximum allowed shared lattice parameter between the core and the shell, i.e., $\epsilon_{||}^{max}=\frac{a_{(In_{0.15}Ga_{0.85}As)}\:-\:a_{(GaAs)}}{a_{(GaAs)}}=0.01$. Therefore, considering the deduced linear relation between the strain and the curvature (figure \ref{fig:6_2}(d)), the maximum predicted curvature that occurs at $\epsilon_{||}^{max}$ will not exceed $\kappa\approx0.25 \: \mathrm{\mu m ^{-1}}$ . For this prediction the associated maximum effective shell growth time is $\approx$ 55 minute that generates the axial strain and nanowire bending. However, this prediction holds only for the given core-shell system and it may differ for different parameters such as nanowire core diameter and In concentration as well as the inhomogeneity degree of the shell distribution.

\noindent To extend on our observations and get a deeper insight into the nanowire bending and strain evolution at the early stages of shell growth, we performed \textit{in-situ} XRD measurement on a single nanowire during shell growth. For this measurement, we scanned a smaller angular range in the vicinity of the GaAs(111) Bragg reflection in order to increase the time resolution to about 11 s. As the Si substrate is expected to be un-strained and straight, its Bragg reflection can be used as a reference to calculate the nanowire bending and the strain within its lattice.  Accordingly, the Si(111) Bragg reflection was subsequently recorded after every 7 scans on GaAs(111) reflection. This explains the gaps in between each 7 scans in figure 7(a) and 7(b). The obtained 2D RSMs and the line profiles extracted along $Q_x$ and $Q_z$ are shown in figure \ref{fig:S4_2}. 

\begin{figure}[!ht]
	\includegraphics[width=\textwidth]{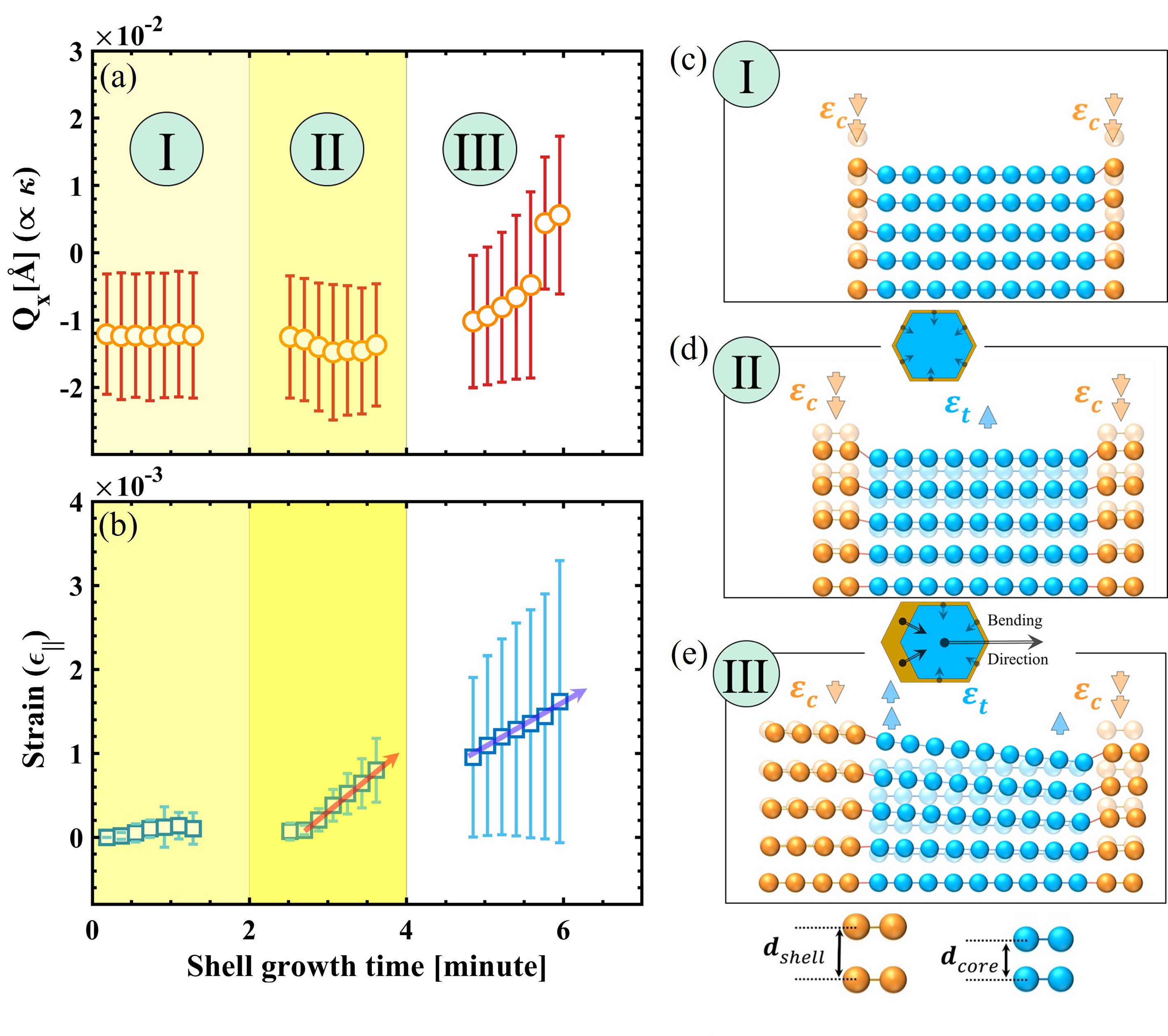}
	\caption{(a) Evolution of the bending angle of the single nanowire during shell growth. (b) Strain evolution in the ZB polytype of the single nanowire during shell growth of sample 5. (c) Animation of the three phases of shell growth indicating the axial strain of the core-shell nanowire and the resulting changes of the lattice plane orientation and the nanowire bending.}
	\label{fig:7_2}
\end{figure}

\noindent The growth was terminated after 11 minutes of shell growth corresponding to shell thickness of $\approx$ 5.5 nm. 
The position of Bragg peak on $Q_x$ which indicates nanowire bending and the induced strain are plotted as functions of shell growth time in figures \ref{fig:7_2}(a) and \ref{fig:7_2}(b), respectively. It can be seen in \ref{fig:7_2}(a) and \ref{fig:7_2}(b) that both quantities pass through three phases during the early stages of shell growth can be explained as follows

\begin{itemize}[noitemsep,nolistsep]
\item At phase-I, corresponding to about 2 minutes of shell growth time, no clear bending and minor strain is observed. At this stage, the first few monolayers of the growing shell (approximated to be with 1 nm in thickness) are subjected to a compressive strain to match the GaAs lattice and thus, these layers can scarcely influence the lattice parameter of the nanowire core as illustrated in figure \ref{fig:7_2}(c).

\item At phase-II, up to 4 minutes of shell growth time, a clear increase of the strain can be observed, while the nanowire bending angle slightly fluctuates. This observation can be explained by a quasi-uniform shell growth around the nanowire core facilitated by the smooth and un-strained nanowire side-facets. In this case the nanowire surface allows the shell material to diffuse freely and nucleate uniformly around the nanowire which results in a symmetric axial strain (uniform strain) that causes no nanowire bending as illustrated in figure \ref{fig:7_2}(d).

\item At phase-III, both bending angle and strain showed simultaneous increase. At this stage, as the strain increases the diffusivity of the shell material decreases accordingly. As a consequence of the decrement of the shell material diffusivity, the growth rate becomes higher on the nanowire facets that face the direct fluxes than the ones on the opposite side. Therefore, a non-uniform shell thickness around the nanowire results in varying strain across the nanowire cross section and thus nanowire bending as illustrated in figure \ref{fig:7_2}(e). In this case, because of the induced asymmetric strain, the nanowire side that is subjected to higher tensile strain becomes more favorable for the continues shell growth than the other side of the nanowire which results in further increment of the growth rate on the tensile strained side. The opposite side of the nanowire, is subjected to two competitive sources of strain, where a tensile strain results from the shell material that may grow even with low growth rate and a compressive strain caused by nanowire bending. This regime may explain the non-linear increment of the nanowire curvature and strain during shell growth as plotted in figure \ref{fig:6_2}(a) and \ref{fig:6_2}(b), respectively. 
\end{itemize}
These assumptions can be supported by the changes of the strain evolution between phase-II and phase-III. It is clear that the increment of the strain is higher at phase-II (indicated by red arrow in figure \ref{fig:7_2}(b)) than phase-III (blue arrow in figure \ref{fig:7_2}(b)) where a certain amount of the strain is released by the nanowire bending.
The strain variation across the nanowire cross-section is represented by the bars in figure \ref{fig:7_2}(b). This variation was estimated from the broadening of the XRD peak along $Q_z$ as it can be seen by the line profile in figure \ref{fig:S4_2}(a).

\newpage

\section*{Summary and Outlook}
As demonstrated in this research, uniform nanowire bending with controllable curvature along a definable bending direction can be achieved by avoiding substrate rotation during MBE growth of a lattice-mismatched shell.
In this work, using Si substrates covered with native and thermal oxide, we observed different preferable bending directions of GaAs-In$_x$Ga$_{1-x}$As core-shell nanowires. The bending direction which implies the preferable nanowire side facets for shell growth showed a clear dependency not only on the arrangement of the material sources, but also on the type of the oxide covering the substrates. For the given arrangement of the effusion cells in our MBE system we observed that the nanowires bend away from the direction of the Ga flux in case of using substrates covered with thermal oxide. On the other hand, nanowires grown on substrates covered with native oxide bend away from the direction of the As flux. This difference in the bending behavior requires further investigation and justification. 
However, by means of sensitive and non-destructive {\textit{in-situ}} XRD measurements of the axial GaAs(111) Bragg reflection, and benefiting from our previous knowledge of the crystal phase distribution along the nanowire growth axis and the bending direction, we obtained a deep insight into the evolution of the induced axial strain and the resulting nanowire bending.\\
The evolution of the strain and nanowire bending showed non-linear dependency as function of shell shell growth time. The non-linearity of these functions was attributed to the changes of the growth dynamics during shell deposition due to the changes of the surface properties of the nanowire.
The performed \textit{in-situ} XRD investigation on a single nanowire revealed that at the early shell growth stages, first the strain increases while the nanowire stays straight which suggests a uniform strain emerge from a homogeneous distribution of the lattice-mismatches shell around the nanowire core. Afterward, as the shell growth proceed, nanowire bending starts to take place alongside with the increasing strain indicating changes in the homogeneity degree of the shell distribution around the nanowire. The induced asymmetric strain results different growth rate at the opposite nanowire facets which in turn increases as the shell growth proceed.

\section*{Acknowledgements}
The authors thank B. Krause, A. Weisshardt for their support at KIT, as well as S. Dehm and the INT for access to the SEM. We are grateful to  M. Vogel and P. auf dem Brinke for using the SEM at the Institute of Materials Engineering at the University of Siegen. We acknowledge DESY (Hamburg, Germany), a member of the Helmholtz Association HGF, for the provision of experimental facilities. Parts of this research were carried out at PETRA III and we would like to thank S. Francoual and D. Reuther for helping setting up the experiment at the Resonant Scattering and Diffraction beamline P09 and A. Khadievas for helping setting up the experiment at the In-situ and Nano X-ray diffraction beamline P23 as well as M. Lippmann for the support in the Clean Room. Other parts of this research were carried out at the DESY Nanolab and we would like to thank T. F. Keller, A. Jeromin and S. Kulkani for using the SEM. Furthermore, we are grateful to O. Krüger and M. Matalla (Ferdinand-Braun-Institut, Berlin) for electron-beam
lithography and to A. Tahraoui, S. Meister and S. Rauwerdink (Paul-Drude-Institut, Berlin) for substrate preparation.\\
This work was funded by BMBF project 05k16PSA with additional support within the framework of project MILAS.

\newpage
\appendix
\begin{center}
\section*{Supplementary materials}
\end{center}

\renewcommand{\thefigure}{S\arabic{figure}}
\begin{wrapfigure}{R}{0.61\textwidth}

	\includegraphics[width=0.51\textwidth]{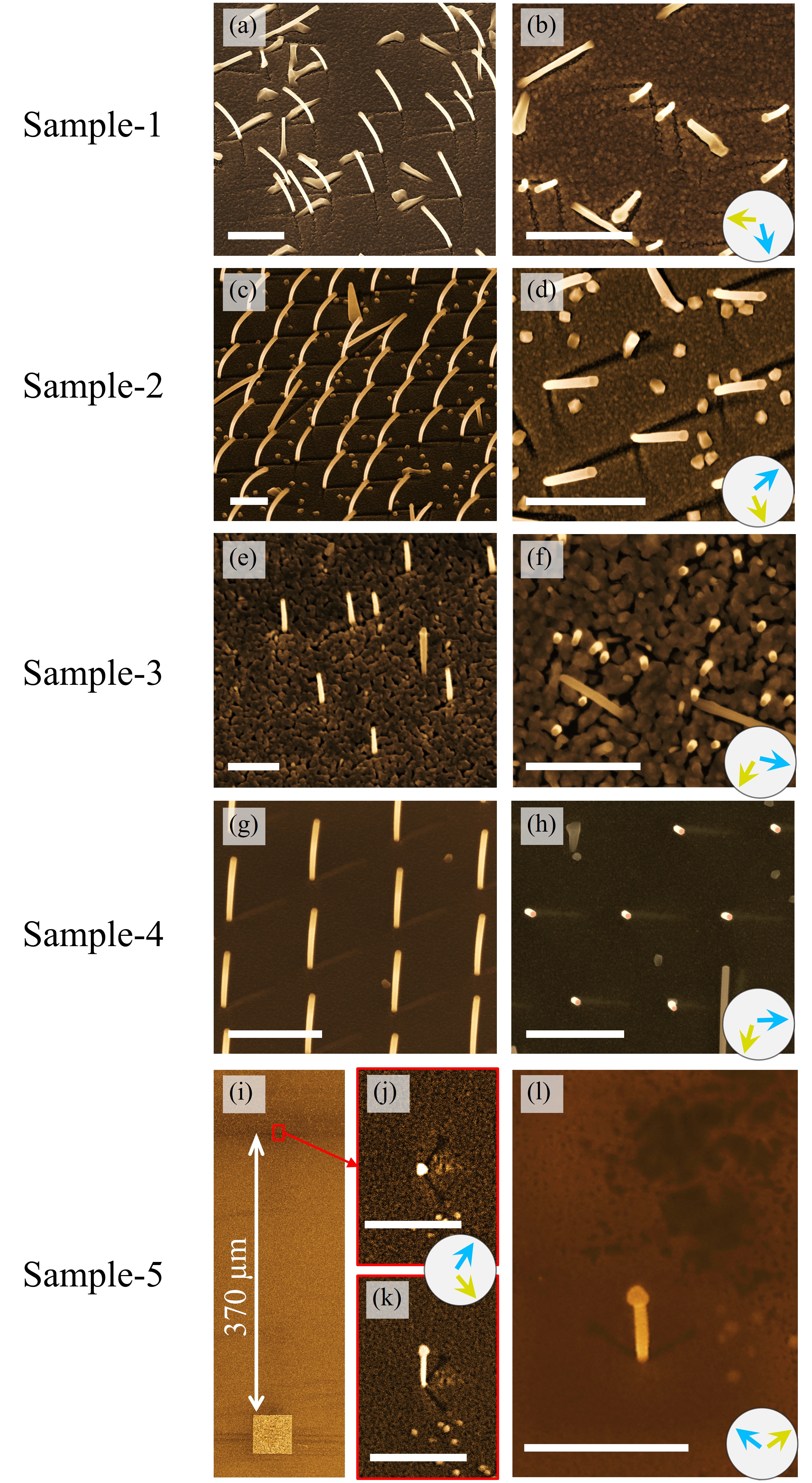}
	\caption {(a)-(h) SEM images of the studied samples with top-view and $30^{\circ}$ tilt-view. (i) Top-view SEM image of sample 5 shows a reference marker and the position of the measured single nanowire. (j) Top view SEM images of the measured single nanowire. (k) and (l) $30^{\circ}$ tilt-view from two different directions of the measured single nanowire. The arrows indicate the direction of Ga flux (blue) and As flux (yellow). All scale bars correspond to 1 $\mathrm{\mu m}$.}
	\label{fig:S1_2}
	\vspace{-110pt} 
\end{wrapfigure}
\newpage
In figure \ref{fig:S1_2}, more SEM images recorded with 30$^{\circ}$ tilt-view and top-view of the studied nanowire samples listed in table \ref{tab1}). 
In this figure, the blue arrows indicate the direction of the Ga flux and yellow arrows indicate the direction of the As flux. As it can be seen from top-view SEM images in figures \ref{fig:S1_2}(b) and \ref{fig:S1_2}(f), the bending direction is in the direction of the As flux for the nanowires grown on native oxide substrates (samples 1 and 3).\\
However, for the nanowires grown on thermal oxide substrates, it can be seen in figures \ref{fig:S1_2}(d), \ref{fig:S1_2}(h) and \ref{fig:S1_2}(l) that the bending direction is in the same direction of the Ga flux. Additionally, figure \ref{fig:S1_2}(i) shows the position of the measured single nanowire on the substrate surface that is located at 370 µm away from the edge of a reference marker. SEM images of the same nanowire are shown with top-view in figure \ref{fig:S1_2}(j) and 30$^{\circ}$ tilt-view from two different azimuths in \ref{fig:S1_2}(k) and \ref{fig:S1_2}(l). \\
\newpage The selected 2D cuts of the RSMs on $Q,Q_z$ component of GaAs(111) Bragg peak of the measured nanowire ensemble (sample 4) are shown in figure \ref{fig:S2_2}(a). These maps are taken at different shell growth times and it can be seen that the Bragg peak is elongating 'horizontally' on a circle (i.e. 2D vertical section of the RSM sphere) toward higher $Q$ values as the shell growth time is increasing (i.e. increasing shell thickness). The elongation along $Q$ indicates the changes of the lattice plane orientation along the nanowire that results from increasing nanowire bending. To evaluate and extract the needed information of these maps, first the WZ signal is integrated on a circle along $Q$ where the radius of this circle is the peak center along $Q_r$. The integrated line profiles of this signal are plotted in figure \ref{fig:2_2}(b) showing that splitting and the slight broadening of the WZ peak is developing as the shell growth time is increasing.
Additionally, shifting and broadening of the XRD signal in RSM toward lower $Q_r$ values can be observed. The vertical shifting indicate increased tensile strain in the GaAs nanowire core and the broadening implies the reduced strain uniformity across the nanowire cross-section while the shell thickness is increasing. In figure \ref{fig:2_2}(c), the line profiles of the ZB phase were extracted along a line on $Q_r$ where this line intersects with $Q$ at the peak center. 
\begin{figure}[!ht]
	\includegraphics[width=\textwidth]{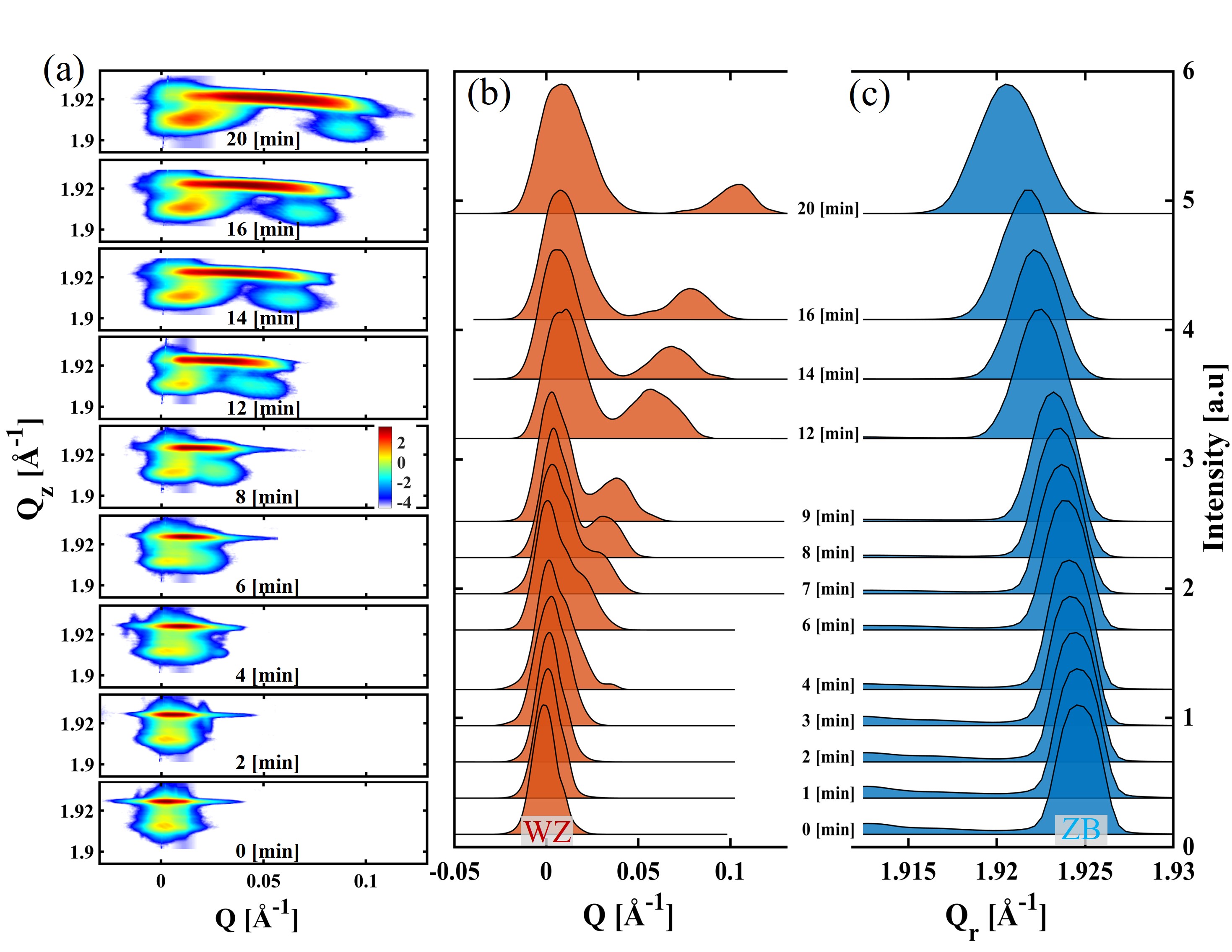}
	\caption {(a) 2D cut in RSM along $Q$ of the GaAs(111) XRD signal on $QQ_z$ of the NW arrays (sample 3) at different shell growth times. (b) The integrated line profiles along $Q$ of the WZ peaks, and (c) along $Q_r$ of the ZB peaks.}
	\label{fig:S2_2}
\end{figure}

\begin{figure}[ht]
    	
    	\includegraphics[width=\textwidth]{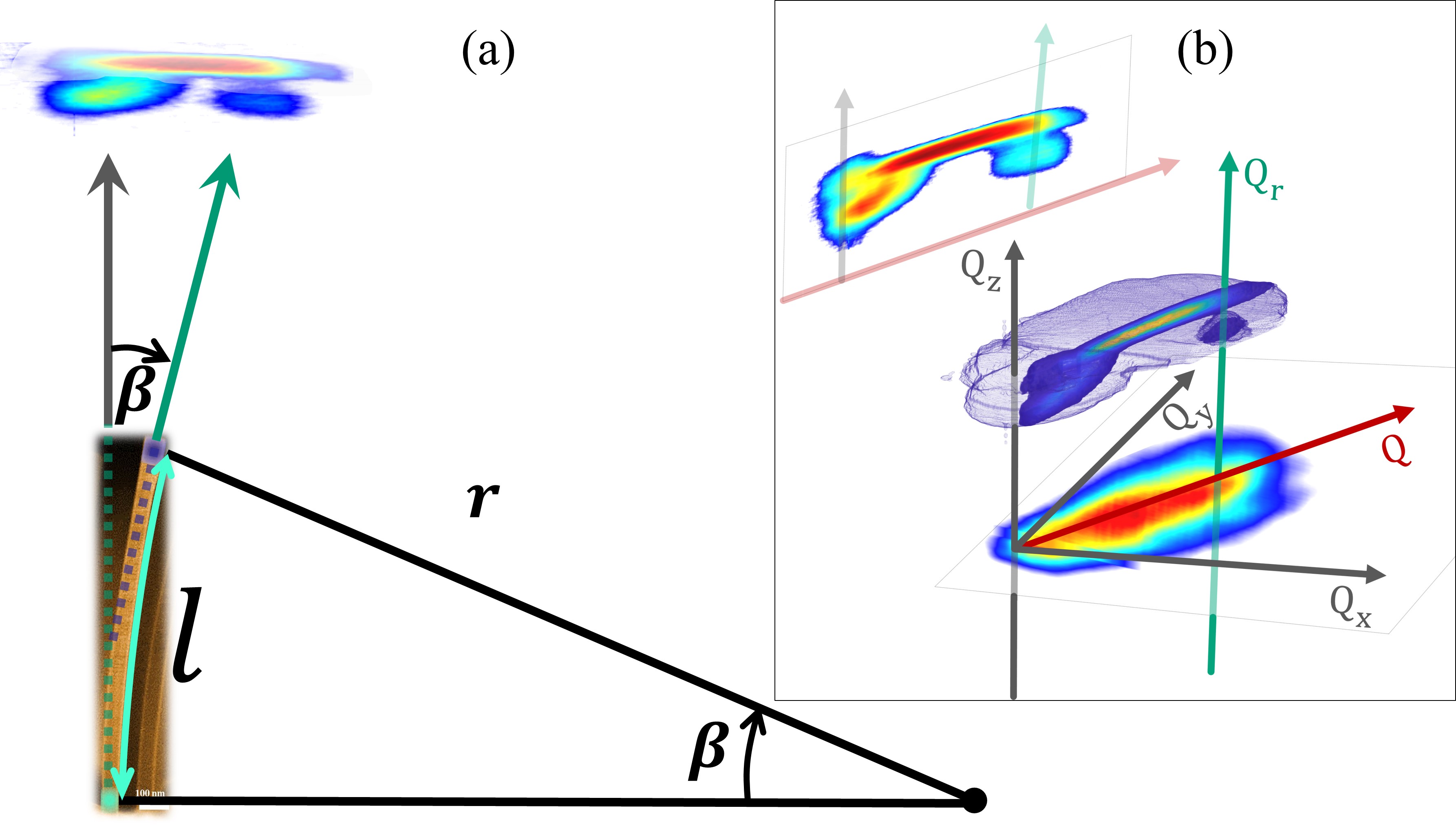}
    	
    	\caption {(a) Illustration of the parameters used for calculating the nanowire curvature combined with side-view SEM image of a bent nanowire in real space and the XRD signal in RSM. (b) A demonstration of the 3D distribution of the XRD signal in RSM and the corresponding RSM vectors, in addition, a 2D cut in RSM of the same signal on $QQ_z$ and a projection on $Q_xQ_y$.}
    	\label{fig:S3_2}
    \end{figure}

\noindent The broadening of the WZ peak at low $Q$ values (i.e. $WZ_{Bottom}$) as well as its higher intensity comparing to $WZ_{Top}$ signal demonstrate the difference in the length of these segments along the nanowire. This observation confirms (in a reversible argument) that nanowire bending may also be used to estimate the axial distribution of ZB and WZ polytypes along the nanowire by XRD measurement. However, for the case of nanowire curvature measurement we consider the position of the $WZ_{top}$ signal in RSM where the bending angle of this segment is the bending angle of the nanowire as illustrated in figure \ref{fig:S3_2}(a).
For the measurement of the nanowire bending angle ($\beta$) we use Gaussian function to fit the XRD peak profile of $WZ_{top}$ and determine its position on $Q$. The peak position on $Q$ at time $t$ ($Q^t(WZ)$) is extracted on the surface of RSM sphere that has a radius $Q^t_r(WZ)$ (see figure \ref{fig:S3_2}). In this case the bending angle is 
\begin{equation}
	\beta=\tan^{-1}\big(\frac{Q^t(WZ)}{Q^t_r(WZ)}\big)
\end{equation}
The tilting angle of the nanowire tip can be used to measure the nanowire curvature $\kappa$ by knowing the average length of the nanowires $l_{NW}$
\begin{equation}
	\centering
	\kappa=\frac{\beta}{l_{NW}}
\end{equation}

\noindent The variation of the peak position along $Q_r$ was used to measure the average strain $\epsilon_{||}$. By considering the position Bragg peak of the unstrained nanowire (i.e. the GaAs nanowire signal before shell growth) as a reference to measure the changes of the lattice parameter of GaAs(ZB) at different shell growth time $t$ as following 

\begin{equation}
	\epsilon_{||}=\frac{a(ZB)^t-a(ZB)^0}{a(ZB)^0}
	\label{EquS3}
\end{equation} 

\noindent where $a(ZB)^0$ and $a(ZB)^t$ are the lattice parameters of GaAs(ZB) before shell growth and at time $t$ of shell growth, respectively. These parameters are measured from the values of the peak center on $Q_r$ at time $t$ which we denoted here as $Q^t_r$
\begin{equation}
	a(ZB)^t=\frac{2\pi}{Q^t_r}
\label{EquS4}
\end{equation} 
Therefore, from equations \ref{EquS3} and \ref{EquS4}, the axial strain is 
\begin{equation}
	\epsilon_{||}=\frac{Q^t_r-Q^0_r}{Q^0_r}
\end{equation}

\noindent The broadening of the peaks along $Q_r$ was considered to measure the strain variation across the nanowire as indicated by bars in figures \ref{fig:6_2}(c) ,\ref{fig:6_2}(d) and \ref{fig:7_2}(b). In this case the standard deviation $\sigma^t_r$ of the signal along $Q_r$ at time $t$ is used to determine the strain variation as follows 

\begin{equation}
	\Delta (Q^t_r)=(Q^t_r-\sigma^t_r)
\end{equation} 
thus
\begin{equation}
	\Delta \epsilon_{||}=\frac{\Delta (Q^t_r)-\Delta (Q^0_r)}{\Delta (Q^0_r)}
\end{equation} 
In this case the peak broadening of XRD signal of the unstrained nanowires (i.e. $\Delta (Q^0_r)$ is taken as a reference value to distinguish contribution of the normal distribution of X-ray the signal that results from the beam profile and the vertical variation of the unstrained nanowires.

\newpage

\begin{figure}[!ht]
	
	\includegraphics[width=\textwidth]{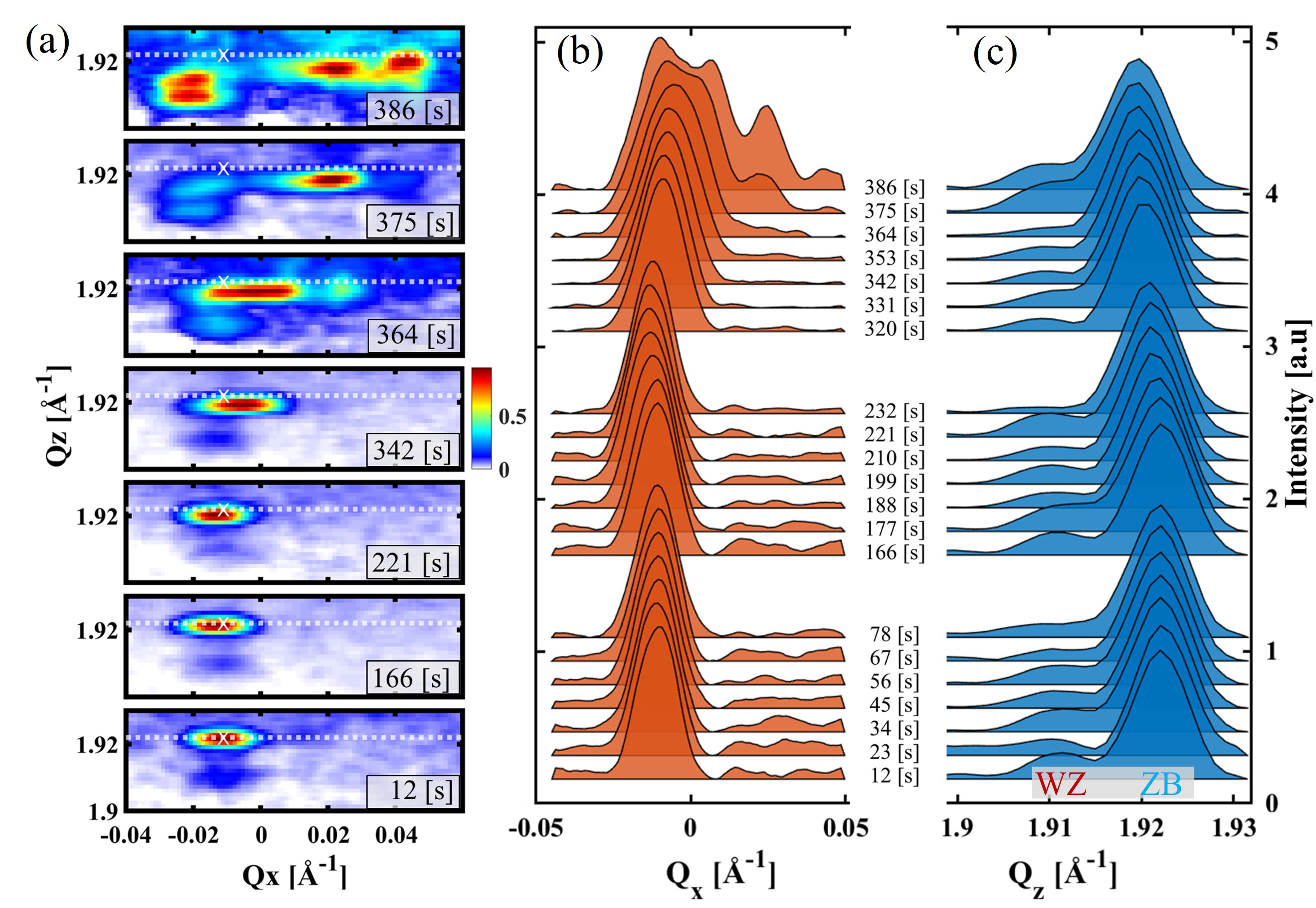}
	\caption {(a) 2D cuts in RSM along $Q_x$ of the GaAs(111) XRD signal on $Q_xQ_z$ of the single nanowire (sample 5) at different shell growth times. (b) The integrated line profiles along $Q_x$ of the entire nanowire signal, and (c) along $Q_r$ of the ZB peaks.}
	\label{fig:S4_2}
\end{figure}

\noindent However, in the case of single nanowire, it was intended to observe the changes of the signal with time resolution of 11 second that is sufficient to obtain enough signal intensity in each frame. Therefore, 2D maps were taken and the data are evaluated on $Q_x$ and $Q_z$ RSM vectors. Few 2D maps taken at different shell growth times are shown in figure \ref{fig:S4_2}(a) and the extracted intensity profiles along $Q_x$ and $Q_z$ of all maps are shown in figures \ref{fig:S4_2}(c) and \ref{fig:S4_2}(d) respectively. After every seven scans, the Si(111) signal were recorded to be used as a reference for the RSM evaluation. 
 
\noindent It can be seen in figure \ref{fig:S4_2}(a) that during the growth of the first few shell layers, the ZB peak slightly shifts toward lower $Q_z$ values (a reference horizontal line is plotted in figure \ref{fig:S4_2}). During further growth, the peak keeps shifting toward lower $Q_z$ values and start to broaden toward higher $Q_x$ values which corresponds to a progressive bending caused by the increasing amount of strain induced by the growing shell.

\newpage


\bibliography{BendingPaper}

\end{document}